\documentclass[prd,aps,showpacs,epsf,floats,onecolumn]{revtex4}
\usepackage[utf8x]{inputenc}
\usepackage{graphicx}
\usepackage{amsfonts}
\usepackage{amsmath}
\usepackage{amssymb}
\usepackage{hyperref}
\usepackage{bookmark}
\usepackage{xcolor}
\usepackage{soul}

\allowdisplaybreaks

\begin{document}

\title{\textbf{A Study of Fluctuation Growths of Observables in Open Quantum Systems}}

\author{\textbf{Newshaw Bahreyni}$^{1}$, \textbf{Paul M. Alsing}$^{2}$, \textbf{Carlo Cafaro}$^{2}$, \textbf{Walid Redjem}$^{2}$, \textbf{Christian Corda}$^{3,4}$} 
\affiliation{$^{1}$Pomona College, Claremont, CA 91711, USA}   
\affiliation{$^{2}$University at Albany-SUNY, Albany, NY 12222, USA}  
\affiliation{$^{3}$SUNY Polytechnic Institute, Utica, NY 13502, USA}
\affiliation{$^{4}$Department of Physics and
Electronics, Christ University, Bengaluru, 560029, Karnataka, India}

\begin{abstract}

The rate at which quantum fluctuations of observables can change provides a useful characterization of the dynamical behavior of quantum systems. For closed quantum systems undergoing unitary evolution, previous work has established a general upper bound on the growth rate of the uncertainty of an observable in terms of the standard deviation of its dynamical velocity.  This result is closely related to quantum acceleration limits and provides a compact constraint on fluctuation growth.

In this work, we will briefly review the closed system results and extend this perspective to open quantum systems, where the state evolves nonunitarily as a consequence of interactions with an environment.  The central result is an upper bound on the rate of growth of the uncertainty of an observable that remains valid for general density-matrix dynamics.  We develop two complementary derivations.  The first introduces a Hermitian pseudo-Hamiltonian describing the local unitary evolution of the eigenvectors of the density matrix.  The second, and more general, derivation proceeds directly from the density-matrix derivative and therefore does not require knowledge of the generator of the dynamics.  Both approaches lead to the same fluctuation-growth bound.  An important practical advantage is that the bound can be evaluated from the observable, the state, and its time derivative without requiring reconstruction of the complete dynamical generator.

\end{abstract}

.

\maketitle

\section{Introduction}

Quantum speed limits (QSLs), which establish fundamental constraints on the rate of quantum evolution, have been extended beyond state evolution to the dynamics of observables and their fluctuations \cite{MP,GP, H1,H2}.  Thus, the dynamical behavior of quantum fluctuations provides important insights into the evolution of quantum systems, and the limits imposed on quantum dynamics.

The standard deviation of an observable quantifies the spread of its possible measurement outcomes and can change as the quantum state evolves or as the observable itself has time dependence.  Understanding how rapidly these fluctuations can grow is therefore  connected to quantum speed limits, quantum acceleration, and open-system evolution \cite{AC,P,CRA,BAC}.

The time evolution of a quantum system can be described in terms of its density operator, which provides a unified framework for investigating both closed and open quantum systems. For a closed system evolving under a possibly time-dependent Hamiltonian $H(t)$, the density matrix $\rho$ satisfies the von Neumann equation (setting $\hbar=1$) \cite{Sa,Coh,vN},
\begin{equation}
\label{eq:0.1}
\frac{\mathrm{d}\rho}{\mathrm{d}t}=-i[H,\rho].
\end{equation}
Such evolution is unitary and preserves the eigenvalues of the density matrix.

The situation is more complicated for open quantum systems, where the density operator generally evolves nonunitarily because of environmental interactions.  
The evolution can be described by a completely positive and trace-preserving (CPTP) map,
\begin{equation}
\label{eq:0.2}
\rho(t)=\mathcal{E}_t[\rho(0)],
\end{equation}
where $\mathcal{E}_t$ represents the dynamical map acting on the initial state \cite{BP}.  For Markovian open quantum systems, the dynamics is commonly described by the Lindblad master equation \cite{Li,M},
\begin{equation}
\label{eq:0.3}
\frac{\mathrm{d}\rho}{\mathrm{d}t}=-i[H,\rho]+\sum_k\left(L_k\rho L_k^\dagger -\frac{1}{2}\{L_k^\dagger L_k,\rho\}
\right),
\end{equation}
where $L_k$ are Lindblad operators describing the coupling between the system and its environment, and $\{X,Y\}=XY+YX$ denotes the anticommutator.  The density-matrix formalism is particularly useful for studying realistic quantum systems, for which environmental interactions can significantly influence the evolution of physical observables.  In this setting, the dynamics of an observable’s variance receive contributions from both the evolution of the quantum state and the explicit time dependence of the observable. Although these contributions can be described through a master equation when the dynamical generator is known \cite{Li,BP}, the underlying Hamiltonian and dissipative operators that are due to the coupling between the system and the environment, may be difficult to determine in realistic physical systems.  This motivates the search for fluctuation-growth bounds that remain applicable without requiring an explicit microscopic description of the system’s evolution.

In this paper, we first review our previous paper in which we developed a general framework for bounding the rate of change of the uncertainty of an observable $A$ for unitary evolutions \cite{CRA}, then extend that to find the evolutions of the uncertainty of observable $A$ in open quantum systems \cite{BAC}.  Our objective is to establish an upper bound on the rate of fluctuation growth that remains applicable without requiring complete knowledge of the specific dynamical generator.  To this end, we consider two cases.  In the first, the generator of evolution is constructed using a time-ordered expansion for a time-dependent Hamiltonian or a Taylor expansion \cite{BCC,LBC}.  In the second, no explicit information about the generator is assumed to be available.  Finally, applying the Cauchy–Schwarz inequality and a standard quadratic inequality yield a bound expressed in terms of the density-matrix derivative, the observable, and its explicit time derivative.  The same result is obtained with and without the knowledge of the evolution of the system which demonstrates that the resulting expression does not require prior knowledge of a Hamiltonian or a Lindblad operator \cite{BAC}.  We also recover the previously established bound for closed quantum systems as a special case of the open-system formulation, connecting our results to the unitary-dynamics analysis presented in Ref. \cite{CRA}.

The remainder of this paper is organized as follows.  In Section II, we review the fluctuation-growth bound for unitary dynamics and its connection to quantum acceleration.  We then introduce a pseudo-Hamiltonian description of the evolution of density-matrix eigenvectors and derive the corresponding fluctuation growth in Section III.  In Section IV, we obtain a generator-independent bound for open quantum systems.  Our concluding remarks including, the physical interpretation, comparison with the unitary bound, and potential experimental relevance, is in Section V.

\section{Bound on Observable Fluctuations  for Unitary Evolutions}

In this section we will include a derivation of the upper bound of fluctuation growth of observable $A$ as shown in Ref. \cite{CRA}.

Consider observable $A(t)$ with its average value defined by 
\begin{equation}
\label{eq:0.4}
\langle A\rangle=\operatorname{Tr}(\rho A)=\sum_j p_j \left\langle \psi_j\right|A\left|\psi_j\right\rangle .
\end{equation}
In a closed system where $A$ evolves unitarily, the time evolution of the expectation value follows \cite{Sa,Eh},
\begin{equation}
\label{eq:0.5}
\frac{\mathrm{d}\langle A\rangle}{\mathrm{d}t}
=\left\langle \partial_t A\right\rangle
+i\langle[H,A]\rangle,
\end{equation}
where $\partial_t A=\partial A/\partial t$ denotes the explicit time derivative of the observable.  According to Eq. (\ref{eq:0.5}) the corresponding velocity-like observable is defined as

\begin{equation}
v_A = \dot{A} = \frac{\partial A}{\partial t}
+i[H,A],
\label{eq:1.1}
\end{equation}
with $\hbar=1$ throughout the paper, and that, in the Heisenberg picture, $\dot{A}=v_A$.  Using the variance of $A$ given by  
\begin{equation}
\sigma_A^2=\left\langle (\Delta A)^{2}\right\rangle =\langle A^2\rangle-\langle A\rangle^2,
\label{eq:1.2}
\end{equation}
where $\Delta A\stackrel{\text{def}}{=}A-\langle A\rangle$, and similarly, $\sigma_{v_A}^2=\left\langle (\Delta v_A)^2\right\rangle =\langle v_A^2\rangle-\langle v_A\rangle^2$, where $\Delta v_A\stackrel{\text{def}}{=}v_A-\langle v_A\rangle$, we can derive an upper bound on the growth of $\sigma_A$ by differentiating Eq.(\ref{eq:1.2}) with respect to time which yields

\begin{equation}
\frac{\mathrm{d}\sigma_A^2}{\mathrm{d}t}
=\frac{\mathrm{d}\langle (\Delta A)^2\rangle}{\mathrm{d}t}=\left\langle \left\lbrace \Delta A, \Delta \dot{A}\right\rbrace \right\rangle .
\label{eq:1.3}
\end{equation}

We can also write the Cauchy–Schwarz inequality for two Hermitian observables $\Delta A$ and $\Delta \dot{A}$ as
\begin{equation}
\left|\left\langle \Delta A\Delta\dot{A}\right\rangle \right|^2
\leq \left\langle (\Delta A)^2\right\rangle 
\left\langle (\Delta\dot{A})^2\right\rangle 
=\sigma_A^2\sigma_{v_A}^2.
\label{eq:1.4}
\end{equation}
Furthermore, noting that
\begin{align}
4\left|\left\langle \Delta A\Delta\dot{A}\right\rangle \right|^2
={}\left|\left\langle \left[ \Delta A,\Delta\dot{A}\right] \right\rangle \right|^2+\left|\left\langle \left\lbrace \Delta A,\Delta\dot{A}\right\rbrace \right\rangle \right|^2,
\label{eq:1.5}
\end{align}
where $[X,Y]=XY-YX$ denotes the commutator, and since the first term on the right-hand side is nonnegative, it follows that
\begin{equation}
\left|\left\langle \left\lbrace \Delta A,\Delta\dot{A}\right\rbrace \right\rangle \right|^2 \leq 4\left|\left\langle \Delta A\Delta\dot{A}\right\rangle \right|^2.
\label{eq:1.5.1}
\end{equation}
Combining this result with Eq. (\ref{eq:1.4}), we obtain
\begin{equation}
\left|\left\langle \left\lbrace \Delta A,\Delta\dot{A}\right\rbrace \right\rangle \right|^2
\leq 4\sigma_A^2\sigma_{v_A}^2.
\label{eq:1.5.2}
\end{equation}
Finally, substituting Eq.(\ref{eq:1.5.2}) into Eq.(\ref{eq:1.3}), dividing by $\sigma_A^2>0$ and taking the square root of both sides yield the fluctuation-growth bound as
\begin{equation}
\left|\frac{d\sigma_A}{dt}\right|
\leq \sigma_{v_A}.
\label{eq:1.6}
\end{equation}

The connection with the quantum acceleration limit found in Ref. \cite{AC} follows immediately by choosing $A=H$.  In this case, the commutator $[H,H]$ in Eq. (\ref{eq:0.5}) and in Eq. (\ref{eq:1.1}) vanishes, and hence

\begin{equation}
v_H=\frac{\partial H}{\partial t}=\dot{H}.
\label{eq:1.6.1}
\end{equation}
Consequently, Eq.(\ref{eq:1.6}) reduces to

\begin{equation}
\left|\frac{\mathrm{d}\sigma_H}{\mathrm{d}t}\right|
\leq \sigma_{\dot{H}}.
\label{eq:1.7}
\end{equation}
Knowing that $a_{H}=\frac{\mathrm{d}v_{H}}{\mathrm{d}t}=\frac{\mathrm{d}\sigma_{H}}{\mathrm{d}t}=\frac{\mathrm{d}\dot{H}}{\mathrm{d}t}=\ddot{H}$ in projective Hilbert space,  Eq. (\ref{eq:1.7}) is precisely the quantum acceleration limit 
\begin{equation}
\left|a_H \right| \leq \sigma_{\dot{H}},
\label{eq:1.8}
\end{equation}
that was introduced in Ref. \cite{AC}.  Furthermore, from Eqs. (\ref{eq:1.6.1},\ref{eq:1.7}), one can see that $v_H=\sigma_H$.  The same result was also obtained in Ref. \cite{P} from another approach.  Thus, the acceleration bound for quantum evolution emerges as a particular case of a more general constraint on the dynamics of observable fluctuations.  Thus the inequality given by Eq. (\ref{eq:1.7}) constrains the rate of change of the energy uncertainty and, consequently, the acceleration of quantum evolution in projective Hilbert space as shown by Eq. (\ref{eq:1.8}).
In other words, Eq. (\ref{eq:1.6}) generalizes the acceleration limit to arbitrary observables.

Next, we will consider open systems and study fluctuation bounds in such systems.

\section{Fluctuation Growth in Open Quantum Systems under known Evolution}

To find the bounds on fluctuations of observable $A$ we will consider two cases: one in which the evolution of the system is known, and one where no information about the evolution of the system is available.  We will consider the latter in the next section.  To study the first case, we start by finding the evolution operator.  

Consider an open quantum system described by the spectral decomposition of its density matrix,

\begin{equation}
\rho(t)=\sum_j p_j(t)
|\psi_j(t)\rangle\langle\psi_j(t)|,
\label{eq:2.1}
\end{equation}
where $p_j(t)$ are the eigenvalues of the density operator and its eigenstates
$\{|\psi_j(t)\rangle\}$ form an orthonormal basis. For an infinitesimal variation of the average value of observable $A(t)$ we have

\begin{align}
d\langle A\rangle
={}&\sum_j dp_j\langle\psi_j|A|\psi_j\rangle+\sum_j p_j\,d\langle\psi_j|A|\psi_j\rangle,
\label{eq:2.2}
\end{align}
where the average value of $A$ is given by
Eq. (\ref{eq:0.4}). 

We assume that the system evolves through a unitary operator $T(t,t+\mathrm{d}t)$ such that it evolves the orthonormal eigenvectors unitarily, while the eigenvalues may evolve nonunitarily due to environmental interactions such that   evolution from an initial time $t$ to a final time $t+\mathrm{d}t$ can be written as \cite{VRF}

\begin{equation}
|\psi_j(t+\mathrm{d}t)\rangle
=T(t,t+\mathrm{d}t)|\psi_j(t)\rangle,
\qquad T^\dagger T=I.
\label{eq:2.4}
\end{equation}
Let $\Omega(t)$ be the Hermitian operator that is the generator of this evolution such that 

\begin{equation}
i\frac{\mathrm{d} T(t)}{\mathrm{d} t}
=\Omega(t)T(t),
\qquad \Omega^\dagger(t)=\Omega(t).
\label{eq:2.5}
\end{equation}
The operator $\Omega$ will be called the
pseudo-Hamiltonian, since it generates the evolution of the density-matrix eigenvectors rather than the complete open-system dynamics.  In general, it differs from the Hamiltonian appearing in the Lindblad master equation given by Eq. (\ref{eq:0.3}) \cite{Li,M}.
Eq. (\ref{eq:2.5}) is similar to the Schr\"{o}dinger equation for operator $T$.  Knowing that for a time-dependent Hamiltonian $H(t)$, the evolution operator $U(t,t_0)$ satisfies the Schr\"odinger equation \cite{Sch} 

\begin{equation}
i\frac{\partial U(t,t_0)}{\partial t}
=H(t)U(t,t_0),
\label{eq:2.7}
\end{equation}
we can use the Taylor expansion for Eq. (\ref{eq:2.7}) for the evolution operator and apply the results to Eq. (\ref{eq:2.5}) to find $T$.  The same method has been used  for numerical simulation of quantum dynamics \cite{BCC,LBC}.  Expanding $U$ and $H$ about $t_0$, with $\Delta t=t-t_0$, gives
\begin{equation}
\begin{aligned}
U(t,t_0)
&=\sum_{k=0}^{\infty}
\frac{(\Delta t)^k}{k!}U^{(k)}(t_0,t_0),\\
H(t)
&=\sum_{k=0}^{\infty}
\frac{(\Delta t)^k}{k!}H^{(k)}(t_0).
\label{eq:2.8}
\end{aligned}
\end{equation}
Knowing that $U^{(0)}(t_0,t_0)=I$ and substituting Eq. (\ref{eq:2.8}) into Eq. (\ref{eq:2.7}) and comparison of equal powers of $\Delta t$ yield

\begin{equation}
iU^{(1)}(t_0,t_0)=H(t_0),
\label{eq:2.9}
\end{equation}
that is,

\begin{equation}
U^{(1)}=-iH(t_0),
\label{eq:2.10}
\end{equation}
and including the second-order expansion gives

\begin{align}
U(t,t_0)
={}&I-iH(t_0)\Delta t + O(\Delta t^2).
\label{eq:2.11}
\end{align}
For an infinitesimal time step we can write,

\begin{equation}
U(t_0+\mathrm{d}t,t_0)
\simeq I-iH(t_0)\mathrm{d}t.
\label{eq:2.12}
\end{equation}
The same result follows from the time-ordered Dyson expansion, which provides a formal solution for finite-time evolution when the Hamiltonian at different times need not commute \cite{PS,NC}:

\begin{align}
U(t,t_0)
&=\mathcal{T}
\exp\left[-i\int_{t_0}^{t}H(t')\,\mathrm{d}t'\right]
\nonumber\\
&=I+\sum_{n=1}^{\infty}(-i)^n
\int_{t_0}^{t}\mathrm{d}t_1
\int_{t_0}^{t_1}\mathrm{d}t_2\cdots
\int_{t_0}^{t_{n-1}} H(t_1)H(t_2)\cdots H(t_n) \mathrm{d}t_n
\label{eq:2.13}
\end{align}
where $t_1\ge t_2\ge\cdots\ge t_n$ and $\mathcal{T}$ is the time ordering operator. In particular,

\begin{align}
\int_{t_0}^{t_0+\Delta t}H(t_1)\,\mathrm{d}t_1 \approx{}\int_{t_0}^{t_0+\Delta t}\left( H(t_0)+\dot{H}(t_0)\left( t_1 - t_0\right) + ...\right) \mathrm{d}t_1 ={}H(t_0)+O(\Delta t^2).
\label{eq:2.14}
\end{align}
The results found in Eqs. (\ref{eq:2.13},\ref{eq:2.14}) reproduce Eq.(\ref{eq:2.12}).  Hence, the Taylor and Dyson methods yield the same local expansion.

Applying the same reasoning to Eq.(\ref{eq:2.5}) gives the infinitesimal eigenvector evolution operator as

\begin{equation}
T(t+\mathrm{d}t,t)\simeq I-i\Omega(t)\mathrm{d}t.
\label{eq:2.15}
\end{equation}
Equations (\ref{eq:2.4}) and (\ref{eq:2.15}) imply
\begin{equation}
\begin{aligned}
\mathrm{d}|\psi_j\rangle
&=-i\Omega|\psi_j\rangle\,\mathrm{d}t,\\
\mathrm{d}\langle\psi_j|
&=i\langle\psi_j|\Omega\,\mathrm{d}t,
\label{eq:2.16}
\end{aligned}
\end{equation}
since $\Omega$ is Hermitian.  We can now find the evolution of the average value of observable $A$ given by Eq. (\ref{eq:2.2}).  Using Eqs.(\ref{eq:0.5}) and (\ref{eq:2.16}), we obtain

\begin{equation}
\frac{\mathrm{d}\langle A\rangle}{\mathrm{d}t}=\sum_j\dot p_j\langle\psi_j|A|\psi_j\rangle +\langle\partial_t A\rangle+i\langle[\Omega,A]\rangle,
\label{eq:2.17}
\end{equation}
where $[\Omega,A]=\Omega A-A\Omega$.

To simplify and write Eq. (\ref{eq:2.17}) in terms of the density operator, differentiate Eq.(\ref{eq:2.1}) and use Eq.(\ref{eq:2.16}) to write

\begin{equation}
\dot\rho
=\sum_j\dot p_j|\psi_j\rangle\langle\psi_j|
+i[\rho,\Omega].
\label{eq:2.18}
\end{equation}
Taking the trace after multiplication by $A$, and using cyclicity of the trace, yield

\begin{align}
\operatorname{Tr}(\dot\rho A)
&=\sum_j\dot p_j\langle\psi_j|A|\psi_j\rangle
+i\operatorname{Tr}([\rho,\Omega]A)
\nonumber\\
&=\sum_j\dot p_j\langle\psi_j|A|\psi_j\rangle
+i\langle[\Omega,A]\rangle.
\label{eq:2.19}
\end{align}
Now, using the result from Eq. (\ref{eq:2.19}), we can rewrite Eq. (\ref{eq:2.17}) as

\begin{equation}
\frac{\mathrm{d}\langle A\rangle}{\mathrm{d}t}
=\operatorname{Tr}(\dot\rho A)
+\langle\partial_t A\rangle,
\label{eq:2.20}
\end{equation}
which agrees with the general identity
$\mathrm{d}\operatorname{Tr}(\rho A)/\mathrm{d}t
=\operatorname{Tr}(\dot\rho A)+\operatorname{Tr}(\rho \dot{A})$.

The result found by Eq. (\ref{eq:2.20}) will be used to find the evolution of the uncertainty of observable $A$.  To find the fluctuations of the uncertainty of observable $A$, consider the variance of a Hermitian observable $A$ given by Eq. (\ref{eq:1.2}).  Upon differentiating Eq. (\ref{eq:1.2}) one obtains 

\begin{equation}
\dot\sigma_A
=\frac{1}{2\sigma_A}
\left[
\frac{\mathrm{d}\langle A^2\rangle}{\mathrm{d}t}
-2\langle A\rangle\frac{\mathrm{d}\langle A\rangle}{\mathrm{d}t}
\right].
\label{eq:3.2}
\end{equation}
Applying Eq. (\ref{eq:2.20}) to $A^2$ gives

\begin{equation}
\frac{\mathrm{d}\langle A^2\rangle}{\mathrm{d}t}
=\operatorname{Tr}(\dot\rho A^2)
+\langle\partial_t(A^2)\rangle.
\label{eq:3.3}
\end{equation}
Substituting the results from Eqs.(\ref{eq:2.20}, \ref{eq:3.3}) into Eq. (\ref{eq:3.2}) and squaring both sides yield
\begin{equation}
\left( \frac{\mathrm{d}\sigma_{A}}{\mathrm{d}t}\right) ^{2} = \frac{1}{4 \sigma^{2}_{A}}\left( \operatorname{Tr}\left(\dot{\rho}\Delta A^{2}\right)  + \left\langle \partial_{t}A^{2}\right\rangle - 2\left\langle A\right\rangle \left\langle \partial_{t}A\right\rangle \right) ^{2}.
\label{eq:3.3.1}
\end{equation}
Now using the symmetrized covariance defined as

\begin{equation}
\operatorname{Cov}(A,\partial_t A)
=\frac{1}{2}\langle\{A,\partial_t A\}\rangle
-\langle A\rangle\langle\partial_t A\rangle,
\label{eq:3.4}
\end{equation}
where $\{X,Y\}=XY+YX$, and knowing that  
\begin{equation}
\left( \frac{\mathrm{d}\sigma_A}{\mathrm{d}t}\right) ^2=\left( \frac{1}{4\sigma_A^2}\right) \left( \frac{\mathrm{d}\sigma_A^2}{\mathrm{d}t}\right)^2,
\label{eq:3.3.2}
\end{equation}
we can use Eq. (\ref{eq:3.3.1}) to write 
\begin{equation}
\frac{\mathrm{d}\sigma_A^2}{\mathrm{d}t}
=\operatorname{Tr}\big(\dot\rho(\Delta A)^2\big)+2\operatorname{Cov}(A,\partial_t A).
\label{eq:3.5}
\end{equation}
The identity given by Eq. (\ref{eq:3.5}) separates the contribution due to the change of the quantum state from that due to the explicit time dependence of the observable.

To find the bounds on the fluctuations of observable $A$, use the Cauchy--Schwarz inequality to write  

\begin{equation}
\left|\operatorname{Tr}
\big(\rho\,\Delta A\,\partial_t A\big)\right|^2 \le \sigma_A^2\langle(\partial_t A)^2\rangle.
\label{eq:3.6}
\end{equation}
Knowing that $\operatorname{Cov}\left( A,\partial_t A\right) = \frac{1}{2}\operatorname{Tr}\left( \rho (\Delta A)\left( \partial_t A\right) +\rho\left( \partial_t A\right) (\Delta A)\right) =  Re \operatorname{Tr}\left( \rho (\Delta A)\partial_t A\right)$, the corresponding bound for the covariance yields \cite{St}

\begin{equation}
|\operatorname{Cov}(A,\partial_t A)|^2
\le \sigma_A^2\langle(\partial_t A)^2\rangle.
\label{eq:3.7}
\end{equation}
Given that for any real $x$ and $y$ we have the inequality $(x+y)^2\le 2(x^2+y^2)$, we can rewrite Eq. (\ref{eq:3.5}) as

\begin{align}
\left(\frac{d\sigma_A^2}{dt}\right)^2
&=\left( \operatorname{Tr}\big(\dot\rho(\Delta A)^2\big)+2\operatorname{Cov}(A,\partial_t A)\right) ^2
\nonumber\\
&\le 2\left( \operatorname{Tr}\big(\dot\rho(\Delta A)^2\big)\right) ^2+8\left( \operatorname{Cov}(A,\partial_t A)\right) ^2
\nonumber\\
&\le 2\left( \operatorname{Tr}\big(\dot\rho(\Delta A)^2\big)\right) ^2+
8\sigma_A^2\langle(\partial_t A)^2\rangle,
\label{eq:3.9}
\end{align}
where Eq.(\ref{eq:3.7}) was used in the last step.
Since $\mathrm{d}\sigma_A^2/\mathrm{d}t=2\sigma_A\dot\sigma_A$, division by
$4\sigma_A^2$, for $\sigma_A>0$, gives

\begin{equation}
\left(\frac{\mathrm{d}\sigma_A}{\mathrm{d}t}\right)^2
\le 2\langle(\partial_t A)^2\rangle
+\frac{\left[\operatorname{Tr}
\big(\dot\rho(\Delta A)^2\big)\right]^2
}{2\sigma_A^2}.
\label{eq:3.10}
\end{equation}
Equation (\ref{eq:3.10}) provides a bound on the instantaneous growth rate of the uncertainty of observable $A$.  The first term characterizes the explicit time dependence of the observable, whereas the second term accounts for changes in the quantum state, including the eigenvector dynamics generated by $\Omega$.  Such fluctuation bounds complement the broader study of quantum speed limits and observable dynamics \cite{H2,GP, CAR1, CAR2}. 

Next, we will study fluctuation bounds for an unknown evolution.

\section{Fluctuation Growth With no Knowledge about the Evolution}

In many open quantum systems, the generator of evolution may be unknown or only partially characterized.  Nevertheless, it may still be possible to describe the time dependence of the density operator $\rho(t)$ and of an observable $A(t)$.  This motivates deriving a bound on the growth of fluctuations that does not require explicit knowledge of the pseudo-Hamiltonian governing the evolution.  

To find the evolution of the variance,  we consider Eq. (\ref{eq:3.3.2}) to find each term.  The first term describes the contribution associated with the evolution of the density operator and the second term accounts for the explicit time dependence of the observable.  Since $(\Delta A)^2=A^2-2A\langle A\rangle+\langle A\rangle^2 I$, its derivative is given by

\begin{equation}
\begin{aligned}
\frac{\mathrm{d}}{\mathrm{d}t}(\Delta A)^2
={}& (\partial_t A)A+A(\partial_t A)    -2(\partial_t A)\langle A\rangle\\
&-2A\frac{\mathrm{d}\langle A\rangle}{\mathrm{d}t}+2\langle A\rangle\frac{\mathrm{d}\langle A\rangle}{\mathrm{d}t}.
\end{aligned}
\label{eq:4.1}
\end{equation}
Taking the trace with $\rho$ and using
$\operatorname{Tr}(\rho A)=\langle A\rangle$ yields

\begin{align}
\operatorname{Tr}\!\left(\rho\frac{\mathrm{d}}{\mathrm{d}t}(\Delta A)^2 \right) = \langle (\partial_t A)A\rangle +\langle A(\partial_t A)\rangle
-2\langle A\rangle\langle\partial_t A\rangle,
\label{eq:4.2}
\end{align}
where in obtaining this expression, the terms proportional to $\mathrm{d}\langle A\rangle/\mathrm{d}t$ cancel.  Moreover, since $A$ and $\partial_t A$ are Hermitian, the first two terms in Eq. (\ref{eq:4.2}) are complex conjugates, that is

\begin{equation}
\langle (\partial_t A)A\rangle = \langle A(\partial_t A)\rangle^*.
\label{eq:4.3}
\end{equation}
Consequently Eq. (\ref{eq:4.2}) becomes,

\begin{equation}
\operatorname{Tr}\!\left(        \rho\frac{\mathrm{d}}{\mathrm{d}t}(\Delta A)^2 \right)=   2\operatorname{Re}\langle A(\partial_t A)\rangle -2\langle A\rangle\langle\partial_t A\rangle.
\label{eq:4.4}
\end{equation}
To express this result in terms of covariance, use the symmetrized
covariance of two observables given by Eq. (\ref{eq:3.4}) to rewrite Eq. (\ref{eq:4.4}) as 

\begin{equation}
\operatorname{Tr}\!\left(\rho\frac{\mathrm{d}}{\mathrm{d}t}(\Delta A)^2 \right)= 2\operatorname{Cov}(A,\partial_t A).
\label{eq:4.5}
\end{equation}
Substitution of this result back into Eq.(\ref{eq:3.3.2}) yields

\begin{equation}
\frac{\mathrm{d}\sigma_A^2}{\mathrm{d}t}
 =\operatorname{Tr}\!\left(\dot{\rho}(\Delta A)^2\right)+2\operatorname{Cov}(A,\partial_t A).
\label{eq:4.7}
\end{equation}
This is the exact identity found in the previous section that was given by Eq. (\ref{eq:3.5}).  However, this identity does not assume a particular equation of motion for $\rho$.  
Following the exact steps shown by Eqs. (\ref{eq:3.6}) through (\ref{eq:3.9}) results in the same bound found by Eq. (\ref{eq:3.10}).
 
Since the first term on the right-hand side of Eq.(\ref{eq:3.10}) depends on the explicit time dependence of the observable, and the second term depends on the density-operator derivative and therefore captures the contribution associated with the evolution of the state, neither term requires explicit knowledge of a pseudo-Hamiltonian or of the detailed generator responsible for the evolution.

\section{Conclusion}
In this work, we investigated the rate of change of observable fluctuations in unitary and open quantum systems.  The  result is a general bound on the growth of the standard deviation of an observable that does not require explicit knowledge of the generator governing the system’s evolution. 

For a density operator $\rho(t)$ and a time-dependent observable $A(t)$, the variance satisfies
$\frac{\mathrm{d}\sigma_A^2}{\mathrm{d}t}=\operatorname{Tr}\!\left[\dot{\rho}(\Delta A)^2\right]+2\operatorname{Cov}(A,\partial_t A)$, where the covariance is symmetrized.  This result is obtained regardless of the knowledge about the system's evolution.  This identity provides a useful separation between changes in the quantum state and explicit changes in the observable with respect to time.  By applying the Cauchy–Schwarz inequality to the covariance and bounding the square of the sum of the two contributions, we derived the bound given by Eq. (\ref{eq:3.10}), for $\sigma_A>0$.  The bound applies to quantum systems regardless of specified Hamiltonian or dissipative generator.

The open-system result becomes especially informative when compared with the corresponding closed-system fluctuation bound given by Eq. (\ref{eq:1.6}) \cite{CRA}.  For unitary dynamics, $\dot A =\partial_t A+i[H,A]$, and the variance satisfies $\sigma_A^2 = \langle(\Delta A)^2\rangle$.  Then under unitary evolution one can show that \cite{BAC}
\begin{equation}
\label{eq:5.1}
\frac{d\sigma_A^2}{dt}=\left\langle \left\lbrace \Delta A, \dot{A}\right\rbrace \right\rangle = 2\operatorname{Cov}(A,\dot A),
\end{equation}
where $\operatorname{Cov}(A,\dot A)$ is the symmetrized covariance given by Eq. (\ref{eq:3.7}).  Then applying Cauchy–Schwarz,$|\operatorname{Cov}(A,\dot A)|
\leq \sigma_A\sigma_{\dot A}$ to Eq. (\ref{eq:5.1}), for $\sigma_A\neq0$, yields
\begin{equation}
\label{eq:5.3}
|\dot\sigma_A| \leq \sigma_{\dot A},
\end{equation}
or equivalently, $\dot\sigma_A^2 \leq \sigma_{\dot A}^2$.
The entire Hamiltonian contribution is contained in the single operator $\dot A
= \partial_t A+i[H,A]$.  Whereas, for an open system, $2\sigma_A\dot\sigma_A=\operatorname{Tr}[\dot\rho(\Delta A)^2] +2\operatorname{Cov}(A,\partial_tA) $, where the first term represents state evolution and the second term represents explicit observable evolution.
This compact structure suggests that the unitary bound is comparatively tighter.
  The derivation of the general bound in open systems is then carried out using $(X+Y)^2 \leq 2(X^2+Y^2)$, with $X=\operatorname{Tr}[\dot\rho(\Delta A)^2]$ and $Y=2\operatorname{Cov}(A,\partial_tA)$.  
This equation suggests another reason why the open-system bound may be generally looser, since the cross term $2XY$ is not included in the upper bound.

Since an important feature of the bound given by Eq. (\ref{eq:3.10}) is that all information relevant to the bound is in $\dot\rho$, regardless of the specific evolution, the result becomes important in situations where the microscopic system-environment interaction is complicated or unknown.  

Some examples can be  applications in quantum metrology where the dynamics of observable fluctuations are relevant to the precision of parameter estimation and to the characterization of measurement sensitivity \cite{GLM,PSM}, quantum thermodynamics in which fluctuation-growth bounds may help characterize the temporal behavior of energy fluctuations and other observables in nonequilibrium processes \cite{EHM}, and in decoherence studies where the density-matrix derivative provides a way to quantify how changes in the quantum state affect observable variances without requiring an explicit decomposition of the dynamics into individual environmental coupling operators \cite{CFR,CEP}.  These connections suggest that the proposed bound could provide a useful analytical tool for investigating fluctuations in systems where the microscopic dynamics are incompletely characterized.  Establishing the quantitative advantages in specific metrological, thermodynamic, and decoherence-related applications remains an important direction for future work.

\end{document}